%% file: hybriddof.tex
\documentclass[sigconf]{acmart}

\AtBeginDocument{%
  \providecommand\BibTeX{{%
    \normalfont B\kern-0.5em{\scshape i\kern-0.25em b}\kern-0.8em\TeX}}}

\setcopyright{rightsretained} 
\copyrightyear{2020} 
\acmYear{2020} 
\acmConference{SIGGRAPH '20 Posters}{August 17, 2020}{Virtual Event, USA}
\acmBooktitle{Special Interest Group on Computer Graphics and Interactive Techniques Conference Posters (SIGGRAPH '20 Posters), August 17, 2020}
\acmDOI{10.1145/3388770.3407426}
\acmISBN{978-1-4503-7973-1/20/08}

\acmSubmissionID{pos\_184}

\begin{document}

\title[Hybrid DoF]{Hybrid DoF: Ray-Traced and Post-Processed Hybrid Depth of Field Effect for Real-Time Rendering}

\author{Tan Yu Wei}
\affiliation{%
 \institution{National University of Singapore}
}
\email{yuwei@u.nus.edu}

\author{Nicholas Chua}
\affiliation{%
 \institution{National University of Singapore}
}
\email{nicholaschuayunzhi@u.nus.edu}

\author{Nathan Biette}
\affiliation{%
 \institution{National University of Singapore}
}
\email{nathan.biette@u.nus.edu}

\author{Anand Bhojan}
\affiliation{%
 \institution{National University of Singapore}
}
\email{banand@comp.nus.edu.sg}

\renewcommand{\shortauthors}{Tan et al.}

\begin{abstract}
	Depth of Field (DoF) in games is usually achieved as a post-process effect by blurring pixels in the sharp rasterized image based on the defined focus plane. This paper describes a novel real-time DoF technique that uses ray tracing with image filtering to achieve more accurate partial occlusion semi-transparencies on edges of blurry foreground geometry. This hybrid rendering technique leverages ray tracing hardware acceleration as well as spatio-temporal reconstruction techniques to achieve interactive frame rates.
\end{abstract}



\begin{teaserfigure}
	\centering
	\includegraphics[width=\linewidth]{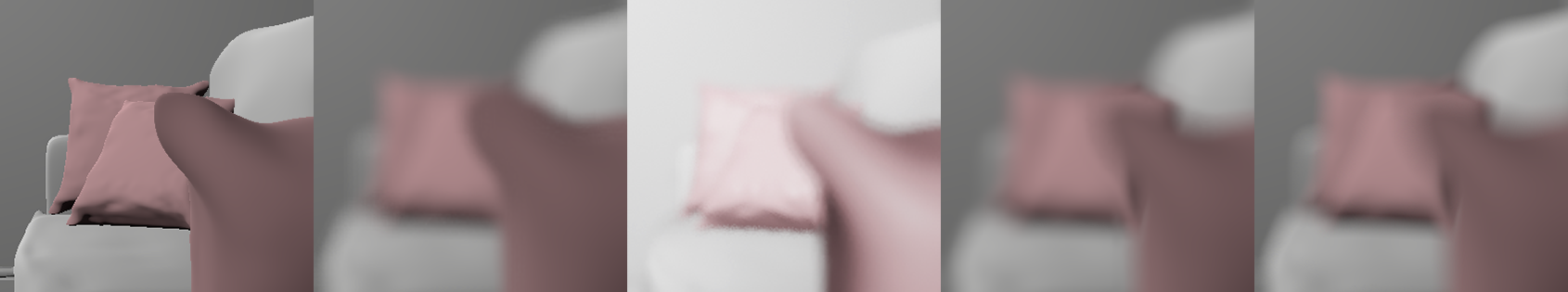}
	\caption{From left to right: original scene, adapted post-process \cite{Jimenez:2014:ARR} with foreground overblur, UE4 post-process \cite{Abadie:2018:ARR}, hybrid (143 fps) and ray-traced DoF on \href{https://www.blendswap.com/blends/view/75692}{\textsc{The Modern Living Room}} (\href{https://creativecommons.org/licenses/by/3.0/}{CC BY}) with a GeForce RTX 2080 Ti}
	\Description{Close-up shot of foreground cushion whose left edge covers bottom right corner of cushion in background}
	\label{fig:teaser}
\end{teaserfigure}

\maketitle
 
\input{introduction}
\input{design}
\input{discussion}

\begin{acks}
	This work is supported by the Singapore Ministry of Education Academic Research grant T1 251RES1812, “Dynamic Hybrid Real-time Rendering with Hardware Accelerated Ray-tracing and Rasterization for Interactive Applications”. 
\end{acks}

\bibliographystyle{ACM-Reference-Format}
\bibliography{hybriddof}

\end{document}

%% file: introduction.tex
\section{Introduction}

Under partial occlusion in Depth of Field (DoF), background information is revealed through the semi-transparent silhouettes of blurry foreground geometry. Partial occlusion is inaccurate with traditional post-processing as the rasterized image does not store any information behind foreground objects. We can query the scene for background intersections with ray tracing but at poorer performance. Hence, we propose a DoF effect based on hybrid rendering, which combines rasterization and ray tracing techniques for better visual quality while maintaining interactive frame rates.

%% file: design.tex
\section{Design}

A G-Buffer is first produced in deferred shading with a sharp rasterized image that undergoes post-process filtering adapted from \citet{Jimenez:2014:ARR}. Selected areas of the scene with higher rates of post-processing inaccuracy are then chosen to undergo distributed ray tracing with spatio-temporal reconstruction. The post-process and ray trace colours are finally composited together. We build our technique on the thin lens model by \citet{Potmesil:1982:SIG}, and define our scene such that points in front of the focus plane are in the \emph{near} field and points behind are in the \emph{far} field. 

\subsection{Ray Trace}

We shoot a variable number of rays into the scene through an adaptive ray mask. Employing a selective rendering approach like in adaptive frameless rendering, we aim to shoot more rays at edges to create clean semi-transparencies but less at regions with fewer details such as plain surfaces. Like Canny Edge Detection, we apply a Gaussian filter on the G-Buffer first to reduce noise and jaggies along diagonal edges. The filtered G-Buffer then passes through a Sobel kernel to approximate the gradient associated with each pixel based on the differences in surface normal and depth with its surrounding pixels, detecting geometry edges for the ray mask. Accounting for temporal variation to reduce noise, we also shoot more rays at areas of high variance in luminance, including newly ray-traced regions. Hence, the ray mask is complemented with a temporally-accumulated variance estimate \cite{Schied:2017:SVF} from the ray trace output of previous frames.

We then shoot rays, taking multiple samples in a circle to produce random ray origins within the lens' area. At the edge of an out-of-focus foreground object, rays originating from the lens will either hit the object or objects behind it, yielding a semi-transparent silhouette. We sort the hits into the \emph{near} and \emph{far} fields. However, to have a smooth colour transition, we split the contribution of pixel colour to each layer based on its Z-distance from the separating focus plane. A hit ratio is also stored, which is the number of rays contributing to the \emph{near} colour divided by the total rays shot. 

\subsection{Reconstruction}

To increase our sample count, we temporally-accumulate \emph{near} and \emph{far} ray trace colours over time and use an exponential moving average to blend between history and current frames like in TAA. However, to account for movement, we leverage per-pixel depth and motion information for reprojection like \citet{Lehtinen:2011:TLF}. 

As such, we use screen space motion vectors to reproject \emph{near} pixels. As for the \emph{far} field, we require an approximation of the \emph{far} world position of our pixel. We first attempt to compute the average world position of our pixel from \emph{far} ray trace hits. Under low ray counts, in the event that there is no \emph{far} hit, we obtain the average \emph{far} world position of neighbouring pixels instead. Then, we use this world position to calculate the previous screen space position of the pixel, approximating reprojection for occluded objects appearing within semi-transparent regions of foreground geometry. However, to prevent the ghosting (or smearing) of these regions, we use the latest hit ratio to compose the \emph{near} and \emph{far} colours during motion, but normalize the final colour based on the accumulated hit ratio since reprojection is different for the \emph{near} and \emph{far} fields. Otherwise, we risk having varying colour intensities in our merged result. 

We also blend the new average \emph{near} and \emph{far} Circle of Confusion (CoC) sizes based on the accumulated hit ratio, so as to get an approximate CoC size for the current frame to scale our circular gathering kernel and determine the level-of-detail for sampling in spatial reconstruction. Similar to our post-processing approach, we gather the colour contributions of neighbouring pixels, and samples with CoC radius smaller than their distance to the pixel are rejected. Finally, the gathered colour is weighted based on clamped variance estimates to avoid over-blurring in converged regions.

\subsection{Composite}

For the final image, we apply the ray trace, post-process and sharp rasterized colours onto pixels based on their Z-values in relation to the focus zone, where the CoC size of pixels is less than $\sqrt{2}$ and hence the unblurred rasterized colour is applied. Outside the focus zone, for \emph{near} objects in general, we apply the ray trace colour to form accurate semi-transparent silhouettes. However, for bright specular bokeh shapes, we favour the post-process over the ray trace colour based on the intensity of the specular component of the post-process colour, so as to minimize noise from the ray trace.

For \emph{far} geometry out of the focus zone, we adaptively blend the ray trace and post-process colours using the hit ratio. At high hit ratios, we favour the post-process colour as fewer rays hit the \emph{far} field. On the other hand, low hit ratios indicate that the number of hits in the \emph{far} field is comparable to that of the \emph{near} field, so more ray trace colour is used. However, there are also fewer \emph{near} hits closer to the edges of the ray mask, favouring the ray trace colour. Hence, using our hit ratio as-is causes blur discontinuities and tiling artifacts from the mask. To minimize them while trying to retain the ray-traced semi-transparencies, we only blend at hit ratios below a certain value for a smooth transition between the ray trace and post-process colours around the edges of the ray mask. 

%% file: discussion.tex
\section{Discussion}

Partial occlusion is more accurate for hybrid DoF as background details which would otherwise be hidden are uncovered in the silhouettes of blurry foreground objects with ray tracing. In contrast, many post-process approaches only perform a local approximation of the background with neighbouring pixels or grow blur out of the silhouette of foreground objects, reusing foreground information to avoid reconstructing the missing background \cite{Jimenez:2014:ARR}. 

In future, we plan to scale the ray mask resolution based on the maximum CoC to prevent tiling artifacts and estimate per-pixel blend factors \cite{Schied:2018:GER} to minimize ghosting. We also intend to extend our approach to support alternative bokeh shapes and not just circular ones of uniform intensities, by varying the shape of our sampling kernel and the relative weight of its samples based on the desired silhouette of the camera aperture and spherical aberration quality of its lens respectively.